\documentclass[10pt,oneside]{article}
\usepackage{amssymb,amsmath}
\usepackage{pstricks, pst-node, pst-tree}
\usepackage[all,poly]{xy}

\newtheorem{theorem}{Theorem}
\newtheorem{lemma}{Lemma}
\newtheorem{corollary}{Corollary}

\newcommand{\qed}{$\square$}
\newcommand{\XX}[1]{%
\Tr{\psset{linestyle=solid}\psframebox{\rule{0pt}{9pt}#1}}}

\newenvironment{proof}{%
  \noindent{\em Proof.\ }}{%
  \hspace*{\fill}\qed
  \vspace{2ex}}

\title{
Compiling Quantum Circuits using the Palindrome Transform
}

\author{
  Alfred V. Aho\thanks{aho@cs.columbia.edu}\\Dept. of Computer Science\\Columbia University\\1214 Amsterdam Avenue\\New York, NY 10027
\and Krysta M. Svore\thanks{kmsvore@cs.columbia.edu}\\Dept. of Computer Science\\Columbia University\\1214 Amsterdam Avenue\\New York, NY 10027}

\begin{document}
\maketitle

\begin{abstract}
The design and optimization of quantum circuits is central to quantum computation.
This paper presents new algorithms for compiling
arbitrary $2^n \times 2^n$ unitary matrices into
efficient circuits of $(n-1)$-controlled single-qubit and $(n-1)$-controlled-NOT gates.
We first present a general algebraic optimization technique,
which we call the Palindrome Transform, that can be used to minimize the number of
self-inverting gates in quantum circuits consisting of concatenations
of palindromic subcircuits.
For a fixed column ordering of two-level decomposition,
we then give an enumerative algorithm for minimal
$(n-1)$-controlled-NOT circuit construction,
which we call the Palindromic Optimization Algorithm.
Our work dramatically reduces the number of gates generated by
the conventional two-level decomposition method for constructing quantum
circuits of $(n-1)$-controlled single-qubit and $(n-1)$-controlled-NOT gates.
\end{abstract}



\section{Introduction}
The recent discovery of algorithms for prime factorization,
discrete logarithms and other important problems \cite{G95, Shor97} that are more efficient
on quantum computers than classical computers
has escalated interest in quantum computing.
However, physical limitations of current quantum technologies, such as coherence time and the number of available qubits,
prevent the usage
of quantum algorithms in any computationally significant setting.
It is important, therefore, for any implementation of a quantum
algorithm to make efficient use of the underlying quantum computing
resources.

No matter what technology will ultimately be used to
implement quantum computers,
the quantum circuit is most likely to remain the primary model
for quantum computation \cite{Deu89,Nielsen,Yao93}.
It allows us to represent an algorithm to be implemented by any quantum
computer as a composition of quantum gates.
Although it is analogous to a classical logic circuit, a 
quantum circuit requires novel compilation and optimization
algorithms since the criteria for efficient quantum computation are radically
different from classical computation. 
It is particularly important to reduce the size of quantum circuits in the 
early phases of compilation 
since the later phases may increase circuit sizes dramatically
for each additional
gate in the initial circuit representation \cite{BBC95,BM02,Kni95,SPM02}.
Ideally we would like to achieve the best circuit for a given class of
gates and a given technology taking into account all
relevant factors such as size, noise, decoherence time, and so forth.
A general-purpose quantum compiler will require both
technology-independent and technology-dependent optimization
techniques to achieve
these efficiency goals.
Until a fully scalable quantum computer technology emerges,
we will restrict ourselves to machine-independent techniques.

In this paper, we focus on the design and optimization of quantum circuits
consisting of controlled single-qubit gates for arbitrary
$2^n \times 2^n$ unitary matrices.
In particular, we focus on the reduction
of $(n-1)$-controlled-NOT gates in such circuits.
To achieve this reduction, we introduce a general algebraic
gate-minimization technique, which we call the Palindrome Transform.
We then present an efficient iterative method,
the Palindromic Optimization Algorithm, for decomposing
a quantum circuit into matrices acting nontrivially on two
or fewer vector components (two-level matrices).
These algorithms are useful in the first phase of any general procedure
for decomposing a quantum computation into an efficient quantum circuit.
Ultimately we would like to produce efficient quantum circuits
for different quantum technologies from high-level specifications
of quantum computations.  

\section{The Quantum Circuit Model} 
We use the standard {\it Dirac} notation for quantum states,
where a quantum state $\psi$ is written in {\it ket} form as $|\psi \rangle$.
A {\it quantum bit}, or {\it qubit} has state $|0\rangle$,
state $|1\rangle$, or a {\it linear combination} of these states,
written as $|\psi\rangle = \alpha|0\rangle + \beta|1\rangle$,
where $\alpha$ and $\beta$ are complex numbers and
$|\alpha|^2 + |\beta|^2 = 1$.
The state space of $n$ qubits, which lie in a $2^n$-dimensional complex Hilbert space,
can be represented as a tensor product of the state space of each single qubit
\begin{eqnarray}\mathbb{C}^2 \otimes \mathbb{C}^2 \otimes \ldots \otimes \mathbb{C}^2 = (\mathbb{C}^2)^{\otimes n} = \mathbb{C}^{2^n}\end{eqnarray}
and a state can be described by the vector
\begin{eqnarray}|\psi\rangle=\sum_{x\in\{0,1\}^n}\alpha_x|x\rangle\end{eqnarray}
where the {\it computational basis states} are of the form
$|x_{n-1}\ldots x_1x_0\rangle$ and the probability of measuring state
$|x\rangle$, where $x=x_{n-1}\ldots x_1x_0$, is $|\alpha_{x}|^2$.

We can model quantum computation using the quantum circuit model
developed by Deutsch \cite{Deu89} and Yao \cite{Yao93}.
The quantum circuit model consists of qubits, quantum wires, and quantum gates,
where quantum wires provide communication between the sequential quantum gates
by transporting output from one computation to serve as input
to another.
To identify the matrix elements of particular quantum gates,
we order our states lexicographically.
In our circuit diagrams, time increases from left to right,
but the order of operators in a matrix sequence is applied
to the state from right to left.  

In the quantum circuit model, a {\it quantum gate} on $n$ qubits is a
$2^n \times 2^n$ unitary matrix $U$.
A composition of quantum gates $G_k \ldots G_1$ is called a {\it quantum circuit} $C$, where the product of $G_k \ldots G_1$ represents the unitary operator computed by $C$. Two quantum circuits are {\it equivalent} if the composition
of their respective gates represents the same unitary matrix.
That is, if circuit $C_1$ represents the matrix $U_1$
and $C_2$ represents $U_2$, and if $U_1=U_2$, then $C_1$ is equivalent to $C_2$.

A set of quantum gates is {\it exactly universal}
if it can represent any unitary operation exactly by a composition of its gates;
a set is {\it approximately universal} if it can approximate
any unitary operation to an arbitrary accuracy by a composition 
of its gates \cite{DBE95,Llo95}.
Since there are noncountably many operations, exact universality
requires an infinite generating set of quantum gates.
However, approximate universality can be achieved by certain discrete sets of quantum gates.
In this paper, we consider exact universality using the universal set
of $(n-1)$-controlled single-qubit and $(n-1)$-controlled-NOT gates \cite{DBE95}.

We use the following standard gates in our quantum circuits.
The single-qubit {\it Pauli-$X$} operator
\begin{eqnarray} X=\left[\begin{array}{cc}0& 1\\1 & 0\end{array}\right] \end{eqnarray}
is similar to the classical NOT operation and takes the state
$|x\rangle \rightarrow |1-x\rangle$. There also exist operations on
multiple qubits, such as the ability to conditionally apply a single-qubit gate.
{\it Control gates} perform the target operation $S$
only if the control qubits are set appropriately.
The $(n-1)$-controlled gate, written as $\Lambda_{n-1}(S)$, denotes $n-1$ qubits
controlling the application of the operator $S$ to the target qubit.
Throughout this paper, $S$ represents a single-qubit gate.
The controlled operation $\Lambda_{n-1}(S)$ is defined by
\begin{eqnarray}\Lambda_{n-1}(S)|x_{n-1}\ldots x_1x_0\rangle|\psi\rangle=|x_{n-1}\ldots x_1x_0\rangle S^{x_{n-1}\wedge\ldots\wedge x_1\wedge x_0}|\psi\rangle\end{eqnarray}
where $x_{n-1}\wedge\ldots\wedge x_1\wedge x_0$ in the exponent of $S$
denotes the Boolean product of the bits $x_{n-1},\ldots,x_1,x_0$.
If the product of these bits is 0, then the operator is not applied.  

The $\Lambda_1(X)$ gate is known as the controlled-NOT gate (CNOT)
and performs the operation $|x,y\rangle \rightarrow |x,x \oplus y\rangle$, where $\oplus$ denotes the logical exclusive-or operation.
Henceforth, we will refer to $\Lambda_1(X)$ as the CNOT gate.
In matrix form, the CNOT gate is
\begin{eqnarray} CNOT=\left[ \begin{array}{cccc}1& 0& 0& 0\\0& 1& 0& 0\\0& 0& 0& 1\\0& 0& 1& 0\end{array}\right] \end{eqnarray}

In this paper, we focus on decomposition techniques using two-level unitary matrices, where a {\it two-level} unitary matrix acts nontrivially on two or
fewer vector components.
Figure \ref{twolevel} shows a two-level matrix $M$.
The row $c$ contains 0's except for the two complex numbers
$\alpha$ and $\beta$ shown.
Likewise the row $r$ contains 0's except for the two complex numbers
$\gamma$ and $\delta$.
The rest of the matrix has 1's on the diagonal and 0's elsewhere.
$M$ acts nontrivially on the space spanned by the row
$c$ and the row $r$.
We define $\tilde{M}$ to be the $2 \times 2$ unitary
submatrix consisting of $\alpha$, $\beta$, $\gamma$ and $\delta$ shown in Figure \ref{comp}.
We call this matrix the {\it component matrix} of $M$.
Clearly, $\tilde{M}$ is a unitary operator that acts on a single qubit.
When necessary, we will indicate the vector components $c$ and $r$ on which $M$ nontrivially acts by writing $M_{c,r}$ and $\tilde{M}_{c,r}$.
\begin{figure}[h] 
\begin{center} 
$M_{c,r}=\left[\begin{array}{cccccccc}1&0&0&0&\cdots&0&0&0\\0&1&0&0&\cdots&0&0&0\\\vdots&\ddots&\vdots&\ddots&\vdots&\ddots&\vdots&\vdots\\0&\cdots&\alpha&\cdots&\beta&\cdots&0&0\\\vdots&\ddots&\vdots&\ddots&\vdots&\ddots&\vdots&\vdots\\0&\cdots&\gamma&\cdots&\delta&\cdots&0&0\\\vdots&\ddots&\vdots&\ddots&\vdots&\ddots&\vdots&\vdots\\0&0&0&0&\cdots&0&0&1\end{array}\right]$
\caption{A generic two-level matrix $M$.}\label{twolevel} 
\end{center} 
\end{figure}
\begin{figure}[h]
\begin{center}
$\tilde{M}_{c,r}=\left[\begin{array}{cc}\alpha&\beta\\\gamma&\delta\end{array}\right]$
\caption{The component matrix $\tilde{M}$ of $M$.}\label{comp}
\end{center}
\end{figure}

\section{A Framework for Quantum Circuit Compilation}
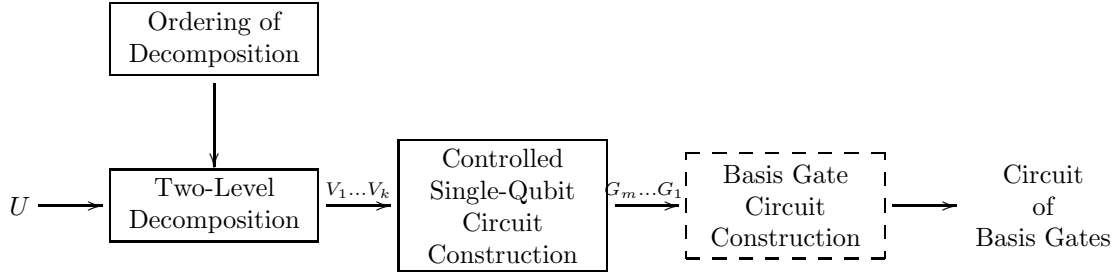
\begin{figure*}
\begin{center}
$\xymatrix{
 & \framebox{\begin{minipage}[c]{1in}\begin{center}Ordering of\\Decomposition\end{center}\end{minipage}} \ar@{->}[d] \\
U \ar@{->}[r] 
& \framebox{\begin{minipage}[c]{1in}\begin{center}Two-Level\\Decomposition\end{center}\end{minipage}} \ar@{->}[r]^{V_1\ldots V_k} 
& \framebox{\begin{minipage}[c]{1in}\begin{center}Controlled\\ Single-Qubit\\ Circuit\\ Construction\end{center}\end{minipage}} \ar@{->}[r]^{G_m\ldots G_1} 
& \dashbox{5}(75,40){\begin{minipage}[c]{1in}\begin{center}Basis Gate\\Circuit\\Construction\end{center}\end{minipage}} \ar@{->}[r]
& \makebox(60,40){\begin{minipage}[c]{1in}\begin{center}Circuit\\of\\Basis Gates\end{center}\end{minipage}}
}$
\caption{The compilation steps of exact quantum circuit generation.}\label{compiler}
\end{center}
\end{figure*}

We now describe the first phase of our quantum circuit compilation process
that generates for an arbitrary unitary matrix $U$ an exact quantum circuit consisting
of $(n-1)$-controlled single-qubit gates and $(n-1)$-controlled-NOT gates \cite{Nielsen, RZBB94}. 
The compilation steps of this phase are shown in Figure \ref{compiler}.

This first phase, called two-level decomposition, takes as input a $2^n \times 2^n$ unitary matrix $U$ and an ordering of decomposition and
outputs a sequence of two-level matrices $V_1\ldots V_k$
such that $V_1\ldots V_k=U$, where $k \leq 2^{n-1}(2^n - 1)$.
This output is then converted into an optimized circuit, $G_m \ldots G_1$, of
$\Lambda_{n-1}(S)$ and $\Lambda_{n-1}(X)$ gates.  Using standard techniques, the circuit of controlled operations can be further decomposed into a circuit 
composed of gates drawn from some universal set of basis gates \cite{BBC95}. One common exactly universal set is the set of single-qubit and CNOT gates \cite{BBC95}.
Our framework here builds on and refines the conventional ordering
and two-level decomposition method
described in \cite{BBC95, Nielsen,RZBB94}.

In this paper, we improve the first phase by finding an optimal ordering of
decomposition for the two-level decomposition phase
to minimize the number of $\Lambda_{n-1}(X)$ gates
generated for the circuit $G_m\ldots G_1$ corresponding to $U$.
The remaining sections of this paper are organized as follows.
In Section 4, we describe the conventional ordering and two-level decomposition
algorithm used in the first step.
In Section 5, we describe the second step that constructs a circuit of controlled single-qubit gates from the sequence of two-level matrices.
In Section 6, we describe the Palindrome Transform that characterizes the
optimal ways to order subcircuits to maximize the amount of cancellation of
self-inverting gates.
In Section 7, we introduce our Palindromic Optimization Algorithm (POA)
that dramatically improves upon the conventional ordering used in the two-level decomposition algorithm of the first phase.
In Sections 8 and 9 we derive equations for the number of generated gates and
compare the sizes of optimized and unoptimized circuits.

\section{Two-Level Decomposition}
We now describe the first phase of our quantum circuit compiler.
This phase, called {\it two-level decomposition},
takes as input an arbitrary $2^n \times 2^n$
unitary matrix $U$ and produces as output a composition of two-level
matrices $V_1\ldots V_k$ such that the product of $V_1\ldots V_k$ equals $U$.
Phase I as described in this section uses the conventional ordering for
two-level decomposition.
In Section 7, we give a method for computing an improved ordering
that dramatically reduces the size of the generated circuit.

We define the {\it order of two-level decomposition} as the
sequence of vector component pairs that are nontrivially acted on by the
two-level matrices in the decomposition $V_1 \ldots V_k$.
We will associate an {\it ordering pair} $(r,c)$ with a two-level matrix $V_j$
to identify the four complex numbers
$V_j[c,c]$, $V_j[c,r]$, $V_j[r,c]$, $V_j[r,r]$ in the component matrix $\tilde{V_j}$.
The sequence of ordering pairs defines the order of the two-level
decomposition.
To avoid repetition in a two-level decomposition, we only allow pairs $(r,c)$
where $r > c$.
Throughout this paper, the first number of an ordering pair represents a row
and the second a column in a matrix.

In all our sequences of ordering pairs, we begin with the pairs for
column 0 followed by those for 1, followed by those for column 2,
and so on up to column $2^n-2$.
We call this a {\it fixed-column} ordering.
In the conventional algorithm for two-level decomposition, the ordering
has the pairs $(c+1,c), (c+2,c),\ldots,(2^n-1,c)$ for column $c$
followed by the pairs $(c+2,c+1),(c+3,c+1),\ldots,(2^n-1,c+1)$
for column $c+1$, and so on.

We will use a triangular array $order_n$ to store the ordering pairs.
The entries in rows $1,2,\ldots,2^n-1-c$ of column $c$ in $order_n$
represent the ordering pairs
$(order_n[1,c],c)$, $(order_n[2,c],c),\ldots, (order_n[2^n-1-c,c],c)$.
For $n=2$, the order array $order_2$ for the conventional algorithm is
\begin{eqnarray*} \left[
\begin{array}{cccc}0& 0& 0& 0\\1& 2& 3& 0\\2& 3& 0& 0\\3& 0& 0& 0\end{array}
\right] \end{eqnarray*}
Note that row $0$ and column $n-1$ are not used in the
two-level decomposition algorithm since they violate the condition
that the row value must be greater than the column value,
but they are included for notational convenience. 
\newline\newline
{\bf Algorithm 1:} Two-Level Decomposition\newline
{\bf Input:} A $2^n \times 2^n$ unitary matrix $U$
and a $2^n\times 2^n$ array $order_n$ dictating the
order of the two-level decomposition.\newline
{\bf Output:} A sequence of two-level matrices $V_1\ldots V_k$
such that $V_1 \ldots V_k = U$.\newline
{\bf Method:}\newline
{\tt procedure} $\mbox{\it TwoLevelDecompose}(U,order_n)$ \{\newline
\indent $M=U$;\newline
\indent $j=1$;\newline
\indent {\tt for} $c=0$ {\tt to} $2^n-2$ {\tt do} \{\newline
\indent\indent {\tt for} $r=order_n[1,c]$ {\tt to} $order_n[2^n-c-1,c]$ {\tt do} \{\newline
\indent\indent\indent {\tt if} $c$ {\tt equals} $2^n-2$ {\tt then} \{\newline
\indent\indent\indent\indent $M_j=I$;\newline
\indent\indent\indent\indent $M_j[c,c]=M[c,c]^*$;\newline
\indent\indent\indent\indent $M_j[c,r]=M[r,c]^*$;\newline
\indent\indent\indent\indent $M_j[r,c]=M[c,r]^*$;\newline
\indent\indent\indent\indent $M_j[r,r]=M[r,r]^*$;\newline
\indent\indent\indent \}\newline
\indent\indent\indent {\tt else if} $M[r,c]$ {\tt equals} $0$ then \{\newline
\indent\indent\indent\indent $M_j=I$;\newline
\indent\indent\indent\indent {\tt if} $r$ {\tt equals} $order_n[2^n-c-1,c]$ {\tt then}\newline
\indent\indent\indent\indent\indent $M_j[c,c]=M[c,c]^*$;\newline
\indent\indent\indent \}\newline
\indent\indent\indent {\tt else} \{\newline
\indent\indent\indent\indent $M_j=I$;\newline
\indent\indent\indent\indent $M_j[c,c]=M[c,c]^*/\sqrt{|M[c,c]|^2+|M[r,c]|^2}$;\newline
\indent\indent\indent\indent $M_j[c,r]=M[r,c]^*/\sqrt{|M[c,c]|^2+|M[r,c]|^2}$;\newline
\indent\indent\indent\indent $M_j[r,c]=M[r,c]/\sqrt{|M[c,c]|^2+|M[r,c]|^2}$;\newline
\indent\indent\indent\indent $M_j[r,r]=-M[c,c]/\sqrt{|M[c,c]|^2+|M[r,c]|^2}$;\newline
\indent\indent\indent \}\newline
\indent\indent\indent $V_j = M_j^\dagger$;\newline
\indent\indent\indent {\tt output} $V_j$;\newline
\indent\indent\indent $M=M_j*M$;\newline
\indent\indent\indent $j=j+1$;\newline
\indent\indent \}\newline
\indent \}\newline
\}

To perform a conventional two-level decomposition on $U$,
we call the procedure $TwoLevelDecompose$ on $U$ and the conventional
ordering array $order_n$ using Algorithm 1.
With the conventional ordering array as input,
the algorithm applies a transformation $M_1$ to $U$ to set
the matrix entry $M_1U[1,0]$ to 0.
It then applies a transformation $M_2$ to $M_1 U$ to set
$M_2M_1U[2,0]$ to 0.
It continues in this fashion until column 0 has a 1 in
the top entry and 0's everywhere else.
This process is sometimes called a {\em quantum Givens operation} \cite{Cyb01}.
It then iteratively applies this process to the $2^n-1 \times 2^n-1$ unitary
submatrix in the lower right-hand corner of $M_{2^n-1} M_{2^n-2} \ldots M_1 U$,
ultimately decomposing $U$ into a product of two-level unitary matrices.

Algorithm 1 produces as output a sequence of two-level unitary matrices
$V_1 \ldots V_k$, where $V_j = M_j^\dagger$, the adjoint of $M_j$.
We can easily verify that $V_1 \ldots V_k = U$,
and that $k\leq 2^{n-1}(2^n-1)$.
We denote the complex conjugate of a complex number $\zeta$
= $a+ib$ as $\zeta^*$ = $a-ib$.

\section{Controlled Single-Qubit Gate Circuit Construction}
After performing the two-level decomposition on $U$,
we need to construct a circuit from the sequence $V_1 \ldots V_k$ of two-level
matrices using $\Lambda_{n-1}(S)$
and $\Lambda_{n-1}(X)$ gates.  
To compute each $V_j$, the circuit must perform a sequence of state
changes in order to bring together the two vector components
that are nontrivially acted on by $V_j$.
The algorithm uses Gray codes to transform each $V_j$ in $V_1 \ldots V_k$
into a circuit of controlled single-qubit gates.
We can determine the state changes needed for $V_j$
by constructing a {\it Gray code}
between the two computational basis states
$|c\rangle$ and $|r\rangle$ of $V_j$.

Let us define $\mbox{\it GrayCode}(c,r)$ between state $|c\rangle$ and state $|r\rangle$
to be a minimal sequence of binary numbers $g_1,g_2,\ldots,g_m$ in which
$g_1$ = $c_{n-1} c_{n-2} \ldots c_{0}$ is the binary expansion of $c$,
$g_m$ = $r_{n-1} r_{n-2} \ldots r_{0}$ is the binary expansion of $r$,
and two adjacent binary expansions $g_j$ and $g_{j+1}$
differ by only one bit for $1\leq j \leq m-1$.
That is, only one bit flip occurs between two binary numbers
in the sequence.
We call the order of bit flips between the binary expansion of
$c$ and the binary expansion of $r$ in the Gray code the
{\it Gray code ordering} for $c$ and $r$.
Note that a bit flip may not be required for every bit position.
Also, there are at most $n+1$ binary numbers in a Gray code
between any pair of states.  From the Gray code sequence, we determine
the corresponding quantum circuit.  

To construct a circuit from the Gray code $g_1,g_2,\ldots,g_m$
for the two-level unitary matrix $V_j$,
we create a $\Lambda_{n-1}(X)$ gate to transform state
$|g_j\rangle$ into $|g_{j+1}\rangle$, for $1\leq j\leq m-2$.
Each gate performs a controlled bit flip
on the differing qubit, conditional that all other qubits
are the same as in states $|g_j\rangle$ and $|g_{j+1}\rangle$.

After the bit-flipping operations,
we create a $\Lambda_{n-1}(\tilde{V_j})$ gate
to transform state $|g_{m-1}\rangle$ into $|g_m\rangle$
with the differing qubit as target and conditional
on all other qubits being the same as in state $|g_m\rangle$.
We then create a sequence of $\Lambda_{n-1}(X)$ gates to
undo the initial sequence of bit-flipping
operations by repeating them in reverse order.

Algorithm 2 presents the details of this circuit-construction process.
It constructs a sequence of controlled single-qubit gates
for each two-level matrix $V_j$ in $V_1 \ldots V_k$.
Note that the output of Algorithm 2 is a sequence of
palindromic subcircuits, subcircuits that read the same
forwards as backwards.
We will discuss the optimization of palindromic circuits
in detail in the next section.

As an example, Table \ref{gray} contains a Gray code between
basis states $|000\rangle$ and $|111\rangle$.
Figure \ref{graycirc} contains the corresponding quantum circuit of five gates,
where $\oplus$ represents the Pauli-$X$ operator,
$\circ$ represents a control on 0, and
$\bullet$ represents a control on 1.
\newline\newline
\begin{table}[h] 
\begin{center} 
\begin{tabular}{|c|c|}\hline
State&Gray Code\\\hline\hline
$|000\rangle$&000\\
   &001\\
   &011\\
$|111\rangle$&111\\\hline
\end{tabular}
\caption{The Gray code between state $|000\rangle$ and state $|111\rangle$.}\label{gray}
\end{center} 
\end{table}
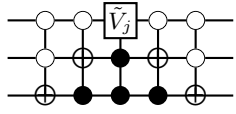
\begin{figure}[h] 
\begin{center}
\begin{pspicture}(-.2,-.5)(3,1.5)
\psline{-}(0,0)(3,0)
\psline{-}(0,.5)(3,.5)
\psline{-}(0,1)(3,1)
\psline{-}(.5,0)(.5,1)
\psline{-}(1,0)(1,1)
\psline{-}(1.5,0)(1.5,1)
\psline{-}(2,0)(2,1)
\psline{-}(2.5,0)(2.5,1)
\psdots[dotstyle=o, dotscale=2](.5,.5)(.5,1)(1,1)(2,1)(2.5,.5)(2.5,1)
\psdots[dotstyle=oplus, dotscale=2.2](.5,0)(1,.5)(2,.5)(2.5,0)
\psdots[dotstyle=*, dotscale=2](1,0)(1.5,0)(1.5,.5)(2,0)
\rput(1.5,1){\psframebox[fillstyle=solid, framesep=1pt]{\small{\ensuremath{\tilde{V}_j}}}}
\end{pspicture}
\caption{The circuit for the two-level matrix $V_j$ that nontrivially acts on states $|000\rangle$ and $|111\rangle$.}\label{graycirc} 
\end{center}
\end{figure}
{\bf Algorithm 2:} Controlled $(n-1)$-Single-Qubit Gate Circuit Construction\newline
{\bf Input:} A sequence of two-level unitary matrices $V_1\ldots V_k$.\newline
{\bf Output:} A circuit composed of
$\Lambda_{n-1}(\tilde{V_j})$
and $\Lambda_{n-1}(X)$ gates, for each $V_j$, $1 \leq j \leq k$,
that computes the product $V_1 \ldots V_k$.\newline
{\bf Method:}\newline
{\tt procedure} $\mbox{\it ConstructCircuit}(V_1 \ldots V_k)$ \{\newline
\indent{\tt for $j=1$ to $k$ do \{\newline
\indent\indent let $|c\rangle$ and $|r\rangle$ be the basis states for $V_j$;\newline
\indent\indent let $g_1,g_2,\ldots,g_m = \mbox{\it GrayCode}(c,r)$;\newline
\indent\indent for $k=1$ to $m-2$ do\newline
\indent\indent\indent output $\mbox{\it ControlGate}(X,g_j,g_{j+1})$;\newline
\indent\indent output $\mbox{\it ControlGate}(\tilde{V_j},g_{m-1},g_m)$;\newline
\indent\indent for $k=m-2$ to $1$ do\newline
\indent\indent\indent output $\mbox{\it ControlGate}(X,g_{j+1},g_j)$;\newline
\indent\}\newline
\}}\newline\newline
{\tt procedure} $\mbox{\it GrayCode}(c,r)$ \{\newline
\indent {\tt let $g$ = $g_{n-1}g_{n-2}\ldots g_0$ be the binary expansion of $c$;\newline
\indent let $h$ = $h_{n-1}h_{n-2}\ldots h_0$ be the binary expansion of $r$;\newline
\indent output $g$;\newline
\indent while $g \neq h$ do \{\newline
\indent\indent let $g_k$ be the rightmost bit in $g$ that is different from\newline
\indent\indent\indent the corresponding bit in $h$;\newline
\indent\indent let $g$ = $g_{n-1} \ldots g_{k+1} \bar{g_k} g_{k-1} \ldots g_0$;\newline
\indent\indent {\bf comment} $\bar{g_k}$ is the complement of $g_k$;\newline
\indent\indent output $g$;\newline
\indent\}\newline
\}}\newline\newline
{\tt procedure} $\mbox{\it ControlGate}(S,g_j,g_{j+1})$ \{\newline 
\indent {\tt output the $(n-1)$-controlled single-qubit gate $\Lambda_{n-1}(S)$\newline
\indent\indent targeting the bit differing between $g_j$ and $g_{j+1}$\newline
\indent\indent and conditional on the other qubits being the same\newline
\indent\indent as in $g_j$;\newline
\}}

\section{The Palindrome Transform}
In this section we present a general algorithmic optimization technique,
which we call the {\it Palindrome Transform},
that can be used to minimize the number of self-inverting
gates in quantum circuits composed of concatenated palindromic subcircuits.
The minimization arises from determining an optimal ordering for
concatenating the palindromic subcircuits that induces the maximal amount
of cancellation due to the juxtaposition of self-inverting gates.
We then characterize the orderings of palindromic subcircuits
that maximize the total amount of cancellation.

We call a gate $A$ {\it self inverting} if $AA = I$, that is,
if $A$ is its own inverse.
If we generate a sequence of
self-inverting gates
of the form
\begin{eqnarray*} A_1 A_2\ldots A_{m-1} A_m A_m A_{m-1}\ldots A_2 A_1\end{eqnarray*}
then we can eliminate this sequence by replacing it with the empty
sequence.
We call such a sequence {\it self annihilating}.

A number of quantum-circuit-generation algorithms produce
subcircuits consisting of sequences of gates in which a prefix and
suffix of each subcircuit forms a palindrome of self-inverting gates.
That is, a subcircuit is of the form
\begin{eqnarray} A_1 A_2\ldots A_k \beta A_k\ldots A_2 A_1\end{eqnarray}
for $m \geq 0$, where each $A_j$ is a self-inverting gate
and $\beta$ is a unique gate that is not necessarily self inverting.
For the purposes of this paper, we assume $\beta$
is a controlled single-qubit gate $\Lambda_{n-1}(S)$,
where $S$ is a component matrix.
We call a sequence of the form (6) a {\it palindromic subcircuit}\footnote{The results in this section also apply to subcircuits of the form
$A_1 \ldots A_k \beta A_k^{-1} \ldots A_1^{-1}$, but these do not arise
in the context of two-level decomposition.}.

If $\alpha$ is a string of symbols $A_1 A_2\ldots A_k$, then we use
$\alpha^R$ to denote $A_k\ldots A_2 A_1$, the reversal of $\alpha$.
Define the {\it overlap} between two palindromic subcircuits
$\alpha_1 A_1 \alpha^R_1$ and $\alpha_2 A_2 \alpha^R_2$
to be the longest reversed suffix $\gamma^R$ of $\alpha^R_1$,
or equivalently the longest prefix $\gamma$ of $\alpha_2$,
such that $\gamma^R \gamma$ is a self-annihilating sequence.

For example, if we concatenate the two palindromic subcircuits
$ABCA_1CBA$ and $ABA_2BA$, we get the circuit
$ABCA_1CBAABA_2BA$ $=$ $ABCA_1CA_2BA$.
Here, $AB$ is the overlap between these two palindromic subcircuits
and $BAAB$ is a self-annihilating sequence.

If we have a set $PS$ of palindromic subcircuits, then we can use
the following algorithm to find an optimal
ordering of all the subcircuits in $PS$ that maximizes the sum of the overlaps
between successive subcircuits
in any composition of the subcircuits.
We call such an ordering a {\it maximal overlap sequence} for $PS$.

The algorithm uses a data structure called a {\it trie} \cite{AHU83},
sometimes called a {\it radix tree} \cite{CLRS01},
to store the prefix $\alpha_j A_j$ of each
palindromic subcircuit $\alpha_j A_j \alpha^R_j$.
The trie is an ordered labeled tree in which there is a path
from the root to a leaf that spells out the string
$\alpha_j A_j$.
The root is labeled by the empty string and each non-root node
is labeled by a gate.
If there is another string $\alpha_k A_k$ that has
a common prefix $\gamma$ with $\alpha_j A_j$,
then the paths for $\alpha_j A_j$ and $\alpha_k A_k$ in the trie
each share the prefix $\gamma$.
For notational convenience, we will just use the middle
$A_j$ to represent a palindromic subcircuit in a maximal overlap sequence.
\newline
\newline
{\bf Algorithm 3:} The Palindrome Transform\newline
{\bf Input:} A set of $m$ palindromic subcircuits
\begin{eqnarray*} PS=\{ \alpha_1 A_1 \alpha^R_1 , \alpha_2 A_2
\alpha^R_2 , \ldots , \alpha_m A_m \alpha^R_m \} \end{eqnarray*}
{\bf Output:} An ordering $A_{j_1},A_{j_2},\ldots,A_{j_m}$
for the concatenation of these palindromic subcircuits such that
\begin{eqnarray*}  \alpha_{j_1} A_{j_1} \alpha^R_{j_1} \alpha_{j_2}
A_{j_2} \alpha^R_{j_2}\ldots \alpha_{j_m} A_{j_m} \alpha^R_{j_m}
\end{eqnarray*}
maximizes
\begin{eqnarray*} \sum_{k=1}^{m-1} length(overlap(\alpha^R_{j_k},
\alpha_{j_{k+1}}))\end{eqnarray*} \newline
where $length(\gamma )$ is the number of gates in the sequence $\gamma$.\newline
{\bf Method:} \newline
{\tt procedure} $\mbox{\it PalindromeTransform}(PS,m)$ \{ \newline
\indent {\tt initialize a trie} $T$; \newline
\indent {\tt for} $j = 1$ {\tt to} $m$ {\tt do} \newline
\indent \indent $\mbox{\it enter}(\alpha_j A_j, T)$; \newline
\indent $\mbox{\it dfsPrint}(T)$; \newline
\} \newline
\newline
{\tt procedure} $\mbox{\it enter}(\mbox{\it string}, T)$ \{ \newline
\indent {\tt let} $\mbox{\it string} = A_1 A_2\ldots A_k$; \newline
\indent {\tt start at root of} $T$; \newline
\indent {\tt follow the longest path} $A_1 A_2 \ldots A_p$ {\tt in $T$ that} \newline
\indent \indent {\tt spells out a prefix of $\mbox{\it string}$ ending at node} $x$; \newline
\indent {\tt create a new path starting at node $x$ that spells out} \newline
\indent \indent $A_{p+1} A_{p+2} \ldots A_k$; \newline
\}
\newline
\newline
{\tt procedure} $\mbox{\it dfsPrint}(T)$ \{ \newline
\indent {\tt visit the nodes of $T$ in a depth-first-search order \newline
\indent \indent printing the label of each leaf when it is first encountered}; \newline
\} \newline
\newline

We call the trie produced by Algorithm 3 the
{\it palindrome trie}.
By entering the $\alpha_j A_j$'s into the trie,
we identify the maximal length common prefixes
for all palindromic subcircuits.
Note that we are using $A_j$ to represent the palindromic
subcircuit $\alpha_j A_j \alpha_j^R$.
By grouping the labels of the leaves of the trie in a
depth-first-search order \cite{AHU83,CLRS01},
we order the palindromic subcircuits to achieve the maximal possible
total overlap of self-inverting gates between successive subcircuits.

\begin{figure}[h]
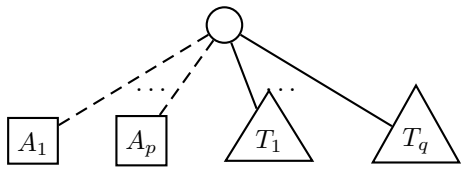

\begin{center}
\pstree[levelsep=*15pt]{\TC}{
\psset{linestyle=dashed}
 \XX{$A_1$}
\trput{$\ldots$}
 \XX{$A_p$}
\psset{linestyle=solid}
 \Ttri{$T_1$}
\trput{$\ldots$}
 \Ttri{$T_q$}
}
\caption{A generic trie.}\label{trie}
\end{center}
\end{figure}

We can characterize the orderings of the leaves of the palindrome
trie that are maximal overlap sequences.
Let $T$ be a trie whose root node has $p$ subtries with exactly one child
labeled $A_1 ,\ldots , A_p$, $p \geq 0$,
and $q$ subtries $T_1 ,\ldots , T_q$, $q \geq 0$,
where each subtrie $T_k$ has more than one child, as shown in Figure \ref{trie}.
We assume that $p+q > 0$ and that the $p+q$ subtries can appear in any order.

Let $mos(T)$ be the set of all sequences of leaf-labels of $T$ that
are characterized by the recurrence
\begin{eqnarray*} mos(T) = \mbox{\it permutation}(A_1 ,\ldots , A_p , mos(T_1),\ldots ,mos(T_q)) \end{eqnarray*}
where $\mbox{\it permutation}(x_1,\ldots ,x_m)$ is the set of all sequences that are
permutations of
the sequences $x_1,\ldots ,x_m$.
We shall show that any sequence in $mos(T)$ is a maximal overlap sequence
and conversely every maximal overlap sequence is in $mos(T)$.
Listing the leaves of the trie in a depth-first-search order 
is one efficient way to produce such a sequence.
\begin{theorem} 
Let $T$ be a palindrome trie for a set $PS$ of palindromic subcircuits.
A sequence of palindromic subcircuits from $PS$ is a maximal overlap
sequence if and only if it is in $mos(T)$.
\end{theorem}
\begin{proof}
To show that every sequence in $mos(T)$ is a maximal overlap sequence
we use structural induction on $T$.
The sequences in $mos(T)$ recursively keep the leaves of the subtries of $T$
contiguous.
Single-leaf subtries of $T$ correspond to palindromic subcircuits
that cannot participate in any prefix sharing.
If $T_j$ is a subtrie of $T$ with $k$ leaves, where $k > 1$,
then $T_j$ adds $2(k-1)$ to the number of cancelling contiguous
self-inverting gates by sharing the gate represented by the branch
from the root of $T$ to the root of the subtrie $T_j$. 
Assuming every sequence in $mos(T_j)$ is a maximal overlap sequence,
then every sequence in $mos(T)$ attains the maximal amount of sharing
and thus maximizes the sum of the lengths of the overlaps between
successive palindromic subcircuits.
Thus every sequence in $mos(T)$ is a maximal overlap sequence.

Conversely, it is easy to show that every maximal overlap sequence
for $PS$ corresponds to some traversal of the palindrome trie for $PS$
represented in $mos(T)$.
\end{proof}
\begin{corollary}
The procedure $\mbox{\it PalindromeTransform}(PS,m)$ produces an ordering for
the $m$ circuits in $PS$ that maximizes the total number of cancelling
self-inverting gates.
\end{corollary}
\begin{proof}
The depth-first-search ordering of the leaves of the palindrome
trie for $PS$ has the $mos$ property.
\end{proof}
\begin{corollary}
The number of gates in the circuit produced by the palindrome transform
ordering after cancelling all self-inverting gates is
\begin{center} $(\mbox{\it number of leaves in trie})+2(\mbox{\it number of interior nodes in trie})$\end{center}
\end{corollary}
\begin{proof}
Note that a path $\alpha_j$ from the root of the palindrome
trie to a leaf labeled by $A_j$ followed by the reverse path
$\alpha_j^R$ defines a palindromic subcircuit $\alpha_j A_j \alpha_j^R$.
One gate is generated for each leaf.
Each incoming branch to an interior node generates one gate before
the leaf to perform an operation and one gate after the leaf to
invert the effect of that operation.
\end{proof}

The palindrome transform assumes the palindromic subcircuits
can be concatenated in any order.
If we treat the middle gate of each palindromic subcircuit
as a generic gate, then we can use the palindrome transform
to generate for an arbitrary unitary matrix $U$ a sequence
of controlled single-qubit gates in which the maximum amount
of cancelling of self-inverting gates takes place,
assuming a fixed column order of two-level decomposition.

To do this, we first construct palindromic subcircuits
with a generic middle gate from the Gray codes for the
conventional ordering of two-level
decomposition for $U$.
From these palindromic subcircuits, we use the palindrome
transform to find an $mos$ ordering of the generic gates.
Using this $mos$ ordering, we then use Algorithms 1 and 2 of the
previous section to construct
the quantum circuit $C$ of $\Lambda_{n-1}(\tilde{V}_j)$ and
$\Lambda_{n-1}(X)$ gates such that $C$ computes $U$.
The circuit $C$ will have the maximal amount of cancellation
of $\Lambda_{n-1}(X)$ gates due to the juxtaposition of
self-annihilating sequences.
Note that any $mos$ ordering produced in this fashion
generates a circuit that computes $U$.

In the next section, we will give a direct enumerative
method of constructing a circuit of this nature without having
to construct the palindrome trie.

\section{\bf Palindromic Optimization Algorithm}
We now describe our Palindromic Optimization Algorithm (POA).
It takes as input a $2^n \times 2^n$ unitary matrix $U$
and produces as output a circuit $G_m \ldots G_1$ of
controlled single-qubit gates that computes $U$
minimizing the number of $\Lambda_{n-1}(X)$ gates
in the generated circuit.

POA performs a two-level decomposition on $U$,
assuming a fixed-column order $0,1,\ldots ,2^n-2$, where
the columns of the matrix are labeled $0$ to $2^n-1$ \cite{Nielsen}.
It uses a specially computed $array_n$ 
to direct the two-level decomposition in order to minimize
the number of $\Lambda_{n-1}(X)$ gates in the generated circuit.
The order of two-level decomposition directs the generation of
a sequence $V_1 \ldots V_k$ of two-level matrices such that $V_1 \ldots V_k=U$.

POA uses Algorithm 2 to generate the output circuit from $V_1 \ldots V_k$.
It uses the Gray code algorithm described in Section 5
to determine the sequences
of $\Lambda_{n-1}(X)$ gates to perform the state changes to
bring together the two nontrivial
vector components for each controlled
$\Lambda_{n-1}(\tilde{V_j})$ gate.
We require the Gray code ordering to be $2^0, 2^1, \ldots , 2^{n-1}$,
where $n$ is the number of qubits, to achieve the minimal number
of $\Lambda_{n-1}(X)$ gates. 
If a different Gray code order is used, the minimal number
of $\Lambda_{n-1}(X)$ gates may not be achieved for all $n$.  
For the stated setting, POA maximizes the overlap of
$\Lambda_{n-1}(X)$ gates over all two-level matrix decompositions,
thus minimizing the
number of $\Lambda_{n-1}(X)$ gates in the generated circuit.
\newline\newline
{\bf Algorithm 4:} Palindromic Optimization Algorithm\newline
{\bf Input:} A $2^n\times 2^n$ unitary matrix $U$ and $n$, the number of qubits.\newline
{\bf Output:} A circuit of $(n-1)$-controlled single-qubit gates that computes $U$.\newline
{\bf Method:}\newline
{\tt procedure} $POA(U)$ \{\newline
\indent $array_n=\mbox{\it ProduceArray}(n)$;\newline
\indent $(V_1\ldots V_k)=\mbox{\it TwoLevelDecompose}(array_n,U)$;\newline
\indent $(G_m\ldots G_1)=\mbox{\it ConstructCircuit}(V_1\ldots V_k)$;\newline
\}\newline\newline
{\tt procedure} $\mbox{\it ProduceArray}(n)$ \{\newline
\indent {\tt $array_2[0..3,0..3] = \left[ \begin{array}{cccc}0& 0& 0& 0\\1& 2& 3& 0\\2& 3& 0& 0\\3& 0& 0& 0\end{array}\right]$;}\newline\newline
\indent {\tt for $m=3$ to $n$ do \{\newline
\indent\indent $k=2^{m-1}$;\newline
\indent\indent for $c=0$ to $2^{m-1}-1$ do \{\newline
\indent\indent\indent $array_m[k,2c] = 2c+1$;\newline
\indent\indent\indent for $r=1$ to $2^{m-1}-c-1$ do \{\newline
\indent\indent\indent\indent $array_m[r,2c]=2array_{m-1}[r,c]$;\newline
\indent\indent\indent\indent $array_m[r+k,2c]=2array_{m-1}[r,c]+1$;\newline
\indent\indent\indent\indent $array_m[r,2c+1]=2array_{m-1}[r,c]$;\newline
\indent\indent\indent\indent $array_m[r+k-1,2c+1]=2array_{m-1}[r,c]+1$;\newline
\indent\indent\indent \}\newline
\indent\indent\indent $k=k-1$;\newline
\indent\indent \}\newline
\indent \}\newline
\indent return $array_m$;\newline
\}}
\begin{figure*}
\begin{center}
\begin{pspicture}(-.2,-.5)(9.5,1)
\psline[linewidth=.1pt]{-}(0,0)(9.5,0)
\psline[linewidth=.1pt]{-}(0,.5)(9.5,.5)
\psline[linewidth=.1pt]{-}(0,1)(9.5,1)
\psline[linewidth=.1pt]{-}(.5,0)(.5,1)
\psline[linewidth=.1pt]{-}(1,0)(1,1)
\psline[linewidth=.1pt]{-}(1.5,0)(1.5,1)
\psline[linewidth=.1pt]{-}(2,0)(2,1)
\psline[linewidth=.1pt]{-}(2.5,0)(2.5,1)
\psline[linewidth=.1pt]{-}(3,0)(3,1)
\psline[linewidth=.1pt]{-}(3.5,0)(3.5,1)
\psline[linewidth=.1pt]{-}(4,0)(4,1)
\psline[linewidth=.1pt]{-}(4.5,0)(4.5,1)
\psline[linewidth=.1pt]{-}(5,0)(5,1)
\psline[linewidth=.1pt]{-}(5.5,0)(5.5,1)
\psline[linewidth=.1pt]{-}(6,0)(6,1)
\psline[linewidth=.1pt]{-}(6.5,0)(6.5,1)
\psline[linewidth=.1pt]{-}(7,0)(7,1)
\psline[linewidth=.1pt]{-}(7.5,0)(7.5,1)
\psline[linewidth=.1pt]{-}(8,0)(8,1)
\psline[linewidth=.1pt]{-}(8.5,0)(8.5,1)
\psline[linewidth=.1pt]{-}(9,0)(9,1)
\psdots[dotstyle=o, dotscale=2](9,.5)(9,1)(8.5,0)(8.5,1)(8,.5)(8,1)(7.5,1)(7,.5)(7,1)(6.5,0)(6.5,.5)(6,.5)(6,1)(5.5,.5)(5,.5)(5,1)(4.5,0)(4.5,1)(4,0)(3.5,0)(3.5,1)(3,.5)(3,1)(2.5,1)(1.5,1)(1,0)(1,1)(.5,.5)(.5,1)
\psdots[dotstyle=oplus, dotscale=2.2](8,0)(7,0)(6,0)(5,0)(4.5,.5)(3.5,.5)(3,0)(2.5,.5)(1.5,.5)(.5,0)
\psdots[dotstyle=*, dotscale=2](7.5,0)(5.5,0)(4,.5)(2.5,0)(2,0)(2,.5)(1.5,0)
\rput(9,0){\psframebox[fillstyle=solid, framesep=1pt]{\small{\ensuremath{\tilde{V}_{1_{0,1}}}}}}
\rput(8.5,.5){{\psframebox[fillstyle=solid,framesep=1pt]{\small{\ensuremath{\tilde{V}_{2_{0,2}}}}}}}
\rput(7.5,.5){\psframebox[fillstyle=solid, framesep=1pt]{\small{\ensuremath{\tilde{V}_{3_{0,3}}}}}}
\rput(6.5,1){\psframebox[fillstyle=solid, framesep=1pt]{\small{\ensuremath{\tilde{V}_{4_{0,4}}}}}}
\rput(5.5,1){\psframebox[fillstyle=solid, framesep=1pt]{\small{\ensuremath{\tilde{V}_{5_{0,5}}}}}}
\rput(4,1){\psframebox[fillstyle=solid, framesep=1pt]{\small{\ensuremath{\tilde{V}_{6_{0,6}}}}}}
\rput(2,1){\psframebox[fillstyle=solid, framesep=1pt]{\small{\ensuremath{\tilde{V}_{7_{0,7}}}}}}
\rput(1,.5){\psframebox[fillstyle=solid, framesep=1pt]{\small{\ensuremath{\tilde{V}_{8_{1,2}}}}}}
\end{pspicture}
\caption{A subsequence of the unoptimized circuit for an arbitrary $2^3 \times 2^3$ unitary matrix using the conventional ordering.}\label{8circ} 
\end{center}
\end{figure*}
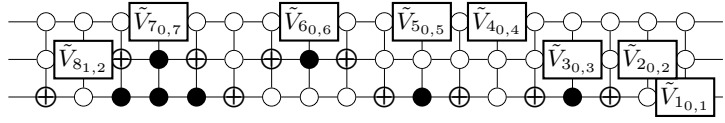
\begin{figure*} 
\begin{center}
\begin{pspicture}(-.2,-.5)(7,1)
\psline[linewidth=.1pt]{-}(0,0)(7,0)
\psline[linewidth=.1pt]{-}(0,.5)(7,.5)
\psline[linewidth=.1pt]{-}(0,1)(7,1)
\psline[linewidth=.1pt]{-}(.5,0)(.5,1)
\psline[linewidth=.1pt]{-}(1,0)(1,1)
\psline[linewidth=.1pt]{-}(1.5,0)(1.5,1)
\psline[linewidth=.1pt]{-}(2,0)(2,1)
\psline[linewidth=.1pt]{-}(2.5,0)(2.5,1)
\psline[linewidth=.1pt]{-}(3,0)(3,1)
\psline[linewidth=.1pt]{-}(3.5,0)(3.5,1)
\psline[linewidth=.1pt]{-}(4,0)(4,1)
\psline[linewidth=.1pt]{-}(4.5,0)(4.5,1)
\psline[linewidth=.1pt]{-}(5,0)(5,1)
\psline[linewidth=.1pt]{-}(5.5,0)(5.5,1)
\psline[linewidth=.1pt]{-}(6,0)(6,1)
\psline[linewidth=.1pt]{-}(6.5,0)(6.5,1)
\psdots[dotstyle=o, dotscale=2](6.5,0)(6.5,1)(6,0)(6,.5)(5.5,0)(5.5,1)(5,0)(4.5,0)(4.5,1)(4,1)(4,.5)(3.5,.5)(3.5,1)(3,1)(2.5,.5)(2,.5)(2,1)(1,1)(.5,0)(.5,1)
\psdots[dotstyle=oplus, dotscale=2.2](5.5,.5)(4.5,.5)(3.5,0)(2,.5)(1,.5)
\psdots[dotstyle=*, dotscale=2](5,.5)(3,0)(2.5,0)(2,0)(1.5,0)(1.5,.5)(1,0)
\rput(4,0){\psframebox[fillstyle=solid, framesep=1pt]{\small{\ensuremath{\tilde{V}_{4_{0,1}}}}}}
\rput(6.5,.5){\psframebox[fillstyle=solid, framesep=1pt]{\small{\ensuremath{\tilde{V}_{1_{0,2}}}}}}
\rput(3,.5){\psframebox[fillstyle=solid, framesep=1pt]{\small{\ensuremath{\tilde{V}_{5_{0,3}}}}}}
\rput(6,1){\psframebox[fillstyle=solid, framesep=1pt]{\small{\ensuremath{\tilde{V}_{2_{0,4}}}}}}
\rput(2.5,1){\psframebox[fillstyle=solid, framesep=1pt]{\small{\ensuremath{\tilde{V}_{6_{0,5}}}}}}
\rput(5,1){\psframebox[fillstyle=solid, framesep=1pt]{\small{\ensuremath{\tilde{V}_{3_{0,6}}}}}}
\rput(1.5,1){\psframebox[fillstyle=solid, framesep=1pt]{\small{\ensuremath{\tilde{V}_{7_{0,7}}}}}}
\rput(.5,.5){\psframebox[fillstyle=solid, framesep=1pt]{\small{\ensuremath{\tilde{V}_{8_{1,2}}}}}}
\end{pspicture}
\caption{A subsequence of the optimized circuit for a $2^3 \times 2^3$ unitary matrix using POA.}\label{Pcircop} 
\end{center}
\end{figure*}
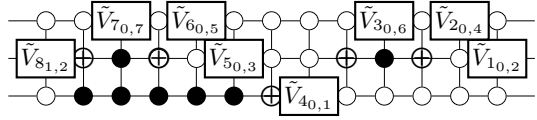
  
We now prove the optimality of POA assuming
a fixed-column ordering $0,1,\ldots,2^n-2$ for a two-level decomposition,
a right-to-left bit ordering $2^0, 2^1,\ldots,2^{n-1}$ for the Gray code order,
and ordering pairs $(r,c)$ in which $r > c$ and the sequence of state changes must occur from $c$ to $r$.

Let $PS(c,r)$ be the palindromic subcircuit generated for the
Gray code sequence returned by the procedure $GrayCode(c,r)$ in Algorithm 2.
First, we examine the intercolumn ordering of the
entries in $array_n$ and the row ordering within a given column
necessary to achieve
a minimal $\Lambda_{n-1}(X)$ circuit for $U$.
Then we prove that the ordering of the entries
from row 1 to row $2^n - c - 1$ in in each column $c$ in $array_n$ is a
maximal overlap sequence for $0 \leq c \leq 2^n - 2$.
\begin{lemma} The maximum possible overlap of $\Lambda_{n-1}(X)$ gates
between the last palindromic subcircuit generated for column $c$
and the first palindromic subcircuit generated for column $c+1$
is 1, for $0 \leq c \leq 2^n - 2$.
Further, an overlap of 1 is achieved between the circuit $PS(c,r_{last})$
followed by the circuit $PS(c+1,r_{first})$, where $r_{last}$
is the last entry in column $c$ and $r_{first}$ is the first
entry in column $c+1$, only when $c$ is even, $r_{last}$ is odd,
and $r_{first}$ is even.
\end{lemma}
\begin{proof}
For $n$ qubits, we have a fixed column ordering $0,1,2,\ldots,2^n-2$.
Let us first consider the case where column $c$ is even.

We would like $PS(c,r_{last})$ and $PS(c+1,r_{first})$ to overlap and
thus share one or more $\Lambda_{n-1}(X)$ gates.
Since $c$ is even and $c+1$ is odd, the $2^0$ bit of the binary expansion
of $c$ is $0$ and the $2^0$ bit of $c+1$ is $1$.
For an overlap to occur, the $2^0$ bit of $r_{last}$ must be $1$
and the $2^0$ bit of $r_{first}$ must be $0$.
Thus, an overlap between subcircuits $PS(c,r_{last})$ and $PS(c+1,r_{first})$ occurs
only when $r_{last}$ is odd and $r_{first}$ is even.
Furthermore, the maximum overlap is 1 since after flipping the $2^0$ bit
of $r_{last}$ to 1, it remains 1.
Similarly, the $2^0$ bit of $r_{first}$ remains 0.
Thus only one overlap can occur.

Now consider the case where column $c$ is odd.
Using the same reasoning as above, an overlap can occur between
$PS(c,r_{last})$ and $PS(c+1,r_{first})$ only when $r_{last}$ is even
and $r_{first}$ is odd.
But, if $c$ is odd, there must be at least one 1 in the binary expansion
of $c+1$ that is not present in $c$.
Since the first bit flip is on bit $2^0$, there cannot
be an overlap due to this differing $1$ and thus the maximum overlap is 0.
\end{proof}
\begin{lemma} Within a column $c$, an overlap can occur between
the subcircuits generated for two adjacent rows only if the entries for both rows
are even or both are odd.
\end{lemma}
\begin{proof} First consider the case where column $c$ is even
and $r_1$ and $r_2$ are the entries for two adjacent rows in column $c$.
We have the following combinations:\newline
i. $r_1$ is odd, $r_2$ is even:  Since only $GrayCode(c,r_1)$
requires a $2^0$ bit flip, $PS(c,r_1)$ and $PS(c,r_2)$
cannot have an overlap.\newline
ii. $r_1,r_2$ are both odd: Since both pairs require a $2^0$ bit flip,
there exists at least one overlap.\newline
iii. $r_1$ is even, $r_2$ is odd: There cannot be an overlap.\newline
iv. $r_1,r_2$ are both even: Since both pairs have a 0 in bit $2^0$,
there may be an overlap.

Similarly, if $c$ is an odd column, then an overlap can occur
only when $r_1$ and $r_2$ are both even or both odd.
\end{proof}

We now prove that POA generates maximal overlap sequences.
Let $R_m^c$ be the sequence
\begin{eqnarray*} array_m[1,c], array_m[2,c],\ldots,array_m[2^m-c-1,c]\end{eqnarray*}
of row entries created by the procedure $ProduceArray$ for column $c$
of $array_m$.
\begin{lemma} 
$R_m^c$ is a maximal overlap sequence, for $0 \leq c \leq 2^m-2$ and
$3 \leq m \leq n$.
\end{lemma}
\begin{proof} We prove by induction on $m$, that $R_m^c$ is a maximal
overlap sequence.
Let the base case be $m=3$.
By inspection of the $2^3\times 2^3$ array $array_3$,
the sequences $R_3^c$ for columns $c = 0,1,\ldots,6$
are maximal overlap sequences.

For the inductive step, assume $R_{m-1}^c$ is a maximal overlap sequence.
Column $c$ of $array_{m-1}$ generates columns $2c$ and $2c+1$ of
$array_m$ as follows:
\begin{align}R_m^{2c}&=2R_{m-1}^c, 2c+1, 2R_{m-1}^c + 1\\
R_m^{2c+1}&=2R_{m-1}^c, 2R_{m-1}^c + 1\end{align}
where
\begin{eqnarray*} 2R_{m-1}^c = 2array_{m-1}[1,c], \ldots, 2array_{m-1}[2^{m-1}-c-1,c]\end{eqnarray*}
and
\begin{eqnarray*} 2R_{m-1}^c+1 = 2array_{m-1}[1,c]+1,\ldots,2array_{m-1}[2^{m-1}-c-1,c]+1.\end{eqnarray*}

Let us now examine how the palindromic subcircuits generated by the
columns of $array_m$ are related to the subcircuits generated from $array_{m-1}$.
Let $PS_m^c$ be the sequence of palindromic subcircuits generated by Algorithm 2
for the row entries in $R_m^c$ in column $c$ of $array_m$.

The GrayCode sequence $\mbox{\it GrayCode}(2c,2r)$ is equivalent to a left shift
of the sequence $\mbox{\it GrayCode}(c,r)$ with a 0 entering in the
$2^0$ bit position in each binary expansion.
Similarly, $\mbox{\it GrayCode}(2c+1,2r+1)$ is equivalent to a left shift
of $\mbox{\it GrayCode}(c,r)$ with a 1 entering in the $2^0$ bit position
in each binary expansion.
Both $\mbox{\it GrayCode}(2c,2r+1)$ and $\mbox{\it GrayCode}(2c+1,2r)$ require one
additional binary expansion in addition to those in $\mbox{\it GrayCode}(c,r)$
since an initial bit flip on bit $2^0$ is now required.

The sequence of palindromic subcircuits $PS_m^{2c}$ is constructed from
the sequence of Gray codes generated by $\mbox{\it GrayCode}(2c,j)$
for all $j$'s in $R_m^{2c}$.
Similarly, the sequence of palindromic subcircuits $PS_m^{2c+1}$ is constructed from
the sequence of Gray codes generated by $\mbox{\it GrayCode}(2c+1,j)$
for all $j$'s in $R_m^{2c+1}$.

We therefore see that the binary code expansions derived from the row entries
in $R_{m-1}^c$ are uniformly shifted.
Further, since $R_m^{2c}$ is the concatentation of $2R_{m-1}^c$
with $2c+1$, $2R_{m-1}^c + 1$, the concatenation does
not generate any new overlaps since $2R_{m-1}$ consists of
even entries, and the entry $2c+1$ and those in $2R_{m-1}^c+1$ are all odd.
Similarly for $R_m^{2c+1}$.
Assuming $R_{m-1}^c$ was a maximal overlap sequence, we conclude
$R_m^{2c}$ and $R_m^{2c+1}$ are also each maximal overlap sequences.
\end{proof}
\begin{theorem} For a fixed-column two-level decomposition of an
arbitrary $2^n \times 2^n$ unitary matrix,
the Palindromic Optimization Algorithm
produces a circuit that achieves the maximal length of overlaps between
successive palindromic subcircuits and thus minimizes the number of
$\Lambda_{n-1}(X)$ gates generated in the quantum circuit of
$(n-1)$-controlled single-qubit and $(n-1)$-controlled-NOT gates.
\end{theorem}
\begin{proof} The proof follows from Lemmas 1-3.
\end{proof}

\section{Gate Count Equations}
We now quantify the number of gates in the circuits
generated by our algorithms.
In all our equations $n$ is the number of qubits.
We first derive the equation for the number of gates produced
by using the conventional two-level decomposition algorithm
assuming no cancelling of self-inverting gates.
We then give the gate count for conventional two-level decomposition
with cancellation.
Finally, we derive the equation that gives the number of gates in
the optimized circuit resulting from performing two-level decomposition
in the order specified by POA.

\subsection{Conventional Circuit Size}
We will show that $c_n$, the number of gates in the unoptimized circuit
produced using the conventional order of two-level decomposition, is given by
\begin{eqnarray} c_n=(n-1)2^{2n-1}+2^{n-1} \end{eqnarray}
We can determine the size of the circuit produced by the
two-level decomposition algorithm for a $2^n \times 2^n$ unitary matrix
using the conventional ordering by taking the number
of Gray codes of length $j$ generated by Algorithm 2, given by
\begin{eqnarray*}2^{n-1} \times {n\choose j}\end{eqnarray*}
and multiplying this number by $2j-1$, the number of gates
in the circuit generated for a Gray code of length $j$.
Thus the number of gates in the conventional circuit for $n$ qubits is
given by
\begin{align*}c_n&=\sum_{j=1}^n {2^{n-1} \times {n\choose j} \times (2j-1)}\\
&= 2^n \times \sum_{j=1}^n (j \times {n\choose j}) - 2^{n-1} \times \sum_{j=1}^n {n\choose j}\\
&= n 2^{2n-1} - 2^{2n-1} + 2^{n-1}\\
&= (n-1) 2^{2n-1} + 2^{n-1}\end{align*}

\subsection{Conventional Circuit Size with Cancelling}
The number of gates in the unoptimized circuit after cancelling
adjacent $\Lambda_{n-1}(X)$ gates between palindromic subcircuits
follows directly from Equation 9.
From Lemmas 1 and 2,
we conclude that only the inter-column overlaps allow for
annihilation of gates using the conventional ordering array for $order_n$.
By Lemma 1, the number of gates that cancel is $2(2^{n-1}-1)$,
so the gate count equation is then
\begin{eqnarray}cc_n=(n-1)2^{2n-1}-2^{n-1}+2\end{eqnarray}

\subsection{POA Circuit Size}
We will show that the number of gates $poa_n$ in the optimal
circuit produced by the Palindromic Optimization Algorithm
for an arbitrary $2^n \times 2^n$ unitary matrix is
\begin{eqnarray} poa_n=(\frac{7}{3})2^{2n-1}-(7)2^{n-1}+\frac{10}{3} \end{eqnarray}

To derive Equation 11 for $2^n\times 2^n$ unitary matrices,
we consider the ordering $array_{n-1}$ and apply POA to determine
$array_n$ and the corresponding number of gates for the circuit for $n$.
From column $c$ of $array_{n-1}$, POA determines columns $2c$ and $2c+1$
of $array_n$.

Consider the case of the even column $2c$ in $array_n$.
We note from the proof of Lemma 3 
that the subtrie for this column is exactly the subtrie for column $c$
in $array_{n-1}$
with two additional branches as given in Equation 7: one branch at one
further depth
containing a copy of the subtrie and a single leaf containing a single gate.
This implies that the number of gates generated by column $2c$ in $array_n$
is twice the number of gates generated column $c$ in $array_{n-1}$ plus
three, two for the additional branch and one for the additional leaf.

Similarly, the odd column $2c+1$ in $array_n$ generates two times
the number of gates generated for column $c$ in $array_{n-1}$
plus two gates required for the additional branch as given in Equation 8.

Note that $R_{n-1}^{2^n-1}$ is empty, so $R_n^{2^n-2}$ contains
a single entry $2^n-1$ and $R_n^{2^n-1}$ is empty.

We can assemble these observations into a recursive formula
to calculate the number of gates in the optimized circuit.
Let $T^c_n$ be the number of gates generated for the $c^{th}$ column of
$array_n$, $0 \leq c \leq 2^n-2$.
We have
\begin{align}T^0_n&=2T^0_{n-1}+3\\
T^1_n&=2T^0_{n-1}+2\\
&\nonumber\vdots\\
T^{2^n-4}_n&=2T^{2^{n-1}-2}_{n-1}+3\\
T^{2^n-3}_n&=2T^{2^{n-1}-2}_{n-1}+2
\end{align}
For the calculation of the two final columns of $array_n$ from the final column of $array_{n-1}$ we have
\begin{align}
T^{2^n-2}_n&=2T^{2^{n-1}-1}_{n-1}+1=1\\
T^{2^n-1}_n&=2T^{2^{n-1}-1}_{n-1}=0
\end{align}

Let $poa_n$ be the total number of gates generated by POA using $array_n$.
Summing the gate counts for every column and recalling that the number
of gates that cancel due to inter-column overlaps is $2(2^{n-1} - 1)$,
$poa_n$ is then given by the recurrence
\begin{eqnarray}poa_n=4(poa_{n-1}+(2^{n-1}-2))+5(2^{n-1}-1)+1-2(2^{n-1}-1)\end{eqnarray}
Solving Equation 18 gives
\begin{eqnarray}poa_n=\sum_{j=n}^{2^n-2}2^j+\sum_{j=1}^{n-1}2^{2j}(2^{n-j}-1)-\sum_{j=1}^{n-2}2^j\label{recur}\end{eqnarray}
Simplifying this equation, we get
\begin{eqnarray*}
poa_n=\frac{7}{3}(2^{2n-1})-7(2^{n-1})+\frac{10}{3}\end{eqnarray*}

\section{Results}
The Palindromic Optimization Algorithm results in a dramatic reduction
in circuit size over the conventional method.
Table \ref{gatenums} lists circuit sizes for $n=2,\ldots, 7$ qubits
resulting from two-level decomposition using the ordering produced by POA, the conventional ordering, and the conventional ordering with no annihilation
of self-inverting gates.

When we use the conventional ordering \cite{Nielsen}
for two-level decomposition on a $2^3 \times 2^3$ unitary matrix,
the resulting circuit contains 62 gates.
Figure \ref{8circ} shows the initial sequence of gates in this circuit.
However, our palindromic optimization algorithm produces a circuit
with 50 gates.
Figure \ref{Pcircop} shows the initial sequence of gates
in this optimized circuit.

The reduction increases linearly with the number of qubits.
For example, when $n = 7$, our method reduces the number of gates
from 49,090 to 18,670 over the conventional method, a more than
$60\%$ reduction. 

\begin{table}[h] 
\begin{center} 
\begin{tabular}{|c|c|c|c|}\hline
$n$&Palindromic&Conventional&No canceling\\\hline\hline
2&8&8&10\\\hline
3&50&62&68\\\hline
4&246&378&392\\\hline
5&1086&2034&2064\\\hline
6&4558&10210&10272\\\hline
7&18670&49090&49216\\\hline
\end{tabular}
\caption{Number of $(n-1)$-controlled gates in an $n$-qubit circuit using our algorithm, the conventional ordering, and the conventional ordering without canceling palindromes.}\label{gatenums}
\end{center} 
\end{table}

\section{Conclusions}
In this paper we have presented a framework for
compiling an arbitrary $2^n \times 2^n$ unitary matrix
into a quantum circuit of $(n-1)$-controlled single-qubit
and $(n-1)$-controlled-NOT gates in which the initial
phase of the framework decomposes
the matrix into a sequence of two-level matrices.
We have shown that the order of two-level decomposition
can have a dramatic impact on the size of the resulting
quantum circuits and we
have characterized those orders of two-level decomposition
that, for a fixed-column ordering, minimize the number
of $(n-1)$-controlled-NOT gates that get generated.
We have also presented an enumerative Palindromic Optimization Algorithm
that produces circuits with the minimal number of
controlled-NOT gates.
This algorithm yields circuits that are significantly
smaller than those produced by the conventional ordering for two-level decomposition.

\section{Acknowledgements}
The authors are grateful to Stephen Edwards and Markus Grassl for many valuable comments and suggestions on the presentation in this paper.


\begin{thebibliography}{99}
\bibitem{AHU83} 
A.~Aho, J.~Hopcroft, and J.~Ullman.
\newblock Data Structures and Algorithms. 
\newblock Addison-Wesley, 1983.

\bibitem{BBC95}
A.~Barenco, C.~Bennett, R.~Cleve, D.~DiVincenzo, N.~Margolus, P.~Shor, T.~Sleator, J.~Smolin, and H.~Weinfurter. 
\newblock Elementary gates for quantum computation.  
\newblock {\em Phys. Rev. A}, 52:3457-3467, 1995. 

\bibitem{BV97} 
E.~Bernstein and U.~Vazirani.  
\newblock Quantum Complexity Theory.
\newblock {\em SIAM J. Comput.}, 26(5):1411-1473, 1997.

\bibitem{BM02}
S. ~Bullock and I. Markov.
\newblock An Arbitrary two-qubit computation in 23 elementary gates.
\newblock quant-ph/0211002, 2003.

\bibitem{CLRS01} 
T.~Cormen, C.~Leiserson, R.~Rivest, and C.~Stein. 
\newblock Introduction to Algorithms, Second Edition. 
\newblock MIT Press, 2001.

\bibitem{Cyb01}
G.~Cybenko.
\newblock Reducing quantum computations to elementary unitary operations.
\newblock {\em Computing in Science and Engineering}, 3(2):27-32, 2001.

\bibitem{DBE95} 
D.~Deutsch, A.~Barenco, and A.~Ekert.  
\newblock Universality in quantum computation.  
\newblock {\em Proc. R. Soc. London A}, 449(1937):669-677, 1995.

\bibitem{Deu89} 
D.~Deutsch. 
\newblock Quantum computational networks.  
\newblock {\em Proc. R. Soc. London A}, 425:73, 1989.

\bibitem{DiV95b} 
D.~DiVincenzo.  
\newblock Two-bit gates are universal for quantum computation.  
\newblock {\em Phys. Rev. A}, 51(2):1015-1022, 1995.

\bibitem{G95}
L.~Grover.
\newblock A fast quantum mechanical algorithm for database search.
\newblock {\em Proc.\ of the 28th Annual Symposium on Theory of Computing}, 1995.

\bibitem{Kni95} 
E.~Knill.  
\newblock Approximating quantum circuits.  
\newblock quant-ph/9905086, 1995.

\bibitem{Llo95} 
S.~Lloyd.
\newblock Almost any quantum logic gate is universal.
\newblock {\em Phys. Rev. Lett.}, 75(2):346, 1995.

\bibitem{Nielsen} 
M.~A.~Nielsen and I.~L.~Chuang. 
\newblock Quantum Computation and Quantum Information. 
\newblock Cambridge University Press, 2000.

\bibitem{RZBB94}
M.~Reck, A.~Zeilinger,H.~J.~Bernstein, and P.~Bertani.
\newblock Experimental realization of any discrete unitary operator.
\newblock {\em Phys. Rev. Lett.}, 73(1):58-61, 1994.

\bibitem{SPM02}
V.~Shende, A.~Prasad, I.~Markov, J.~Hayes.
\newblock Synthesis of reversible logic circuits.
\newblock {\em IEEE Trans. on Computer-Aided Design of Electronic Circuits}, p.714, June 2003.

\bibitem{Shor97}
P.~Shor.
\newblock Polynomial time algorithms for prime factorization and discrete logarithms on a quantum computer.
\newblock {\em SIAM J. of Comput.}, 26(5):1484-1509, 1997.

\bibitem{Yao93}
A.~Yao. 
\newblock Quantum circuit complexity. 
\newblock In {\em Proc.\ of the 34th IEEE Symposium on Foundations of Computer Science}, 1993.

\end{thebibliography}
\end{document}